# Predictive Systems Toxicology

**Narsis A. Kiani**[*124] , **Ming-Mei Shang**[14] , **Hector Zenil**[124], **Jesper Tegner**[*134]

[1] Unit of Computational Medicine, Center for Molecular Medicine, Department of Medicine, Solna, Karolinska Institutet, Stockholm, 171 76, Sweden

[2] Algorithmic Dynamics Lab, Center for Molecular Medicine, Karolinska Institutet, Stockholm, 171 76, Sweden

[3] Biological and Environmental Sciences and Engineering Division, Computer, Electrical and Mathematical Sciences and Engineering Division, King Abdullah University of Science and Technology (KAUST), Thuwal 23955–6900, Kingdom of Saudi Arabia

[4] Science for Life Laboratory, 171 21 Solna, Sweden

**\*Correspondence should be addressed to** narsis.kiani @ki.se and Jesper.Tegner @ki.se

**ABSTRACT:**

In this review we address to what extent computational techniques can augment our ability to predict toxicity. The first section provides a brief history of empirical observations on toxicity dating back to the dawn of Sumerian civilization. Interestingly, the concept of dose emerged very early on, leading up to the modern emphasis on kinetic properties, which in turn encodes the insight that toxicity is not solely a property of a compound but instead depends on the interaction with the host organism. The next logical step is the current conception of evaluating drugs from a personalized medicine point-of-view. We review recent work on integrating what could be referred to as classical pharmacokinetic analysis with emerging systems biology approaches incorporating multiple omics data. These systems approaches employ advanced statistical analytical data processing complemented with machine learning techniques and use both pharmacokinetic and omics data. We find that such integrated approaches not only provide improved predictions of toxicity but also enable mechanistic interpretations of the molecular mechanisms underpinning toxicity and drug resistance. We conclude the chapter by discussing some of the main challenges, such as how to balance the inherent tension between the predicitive capacity of models, which in practice amounts to constraining the number of features in the models versus allowing for rich mechanistic interpretability, i.e. equipping models with numerous molecular features. This challenge also requires patient-specific predictions on toxicity, which in turn requires proper stratification of patients as regards how they respond, with or without adverse toxic effects. In summary, the transformation of the ancient concept of dose is currently successfully operationalized using rich integrative data encoded in patient-specific models.

# A brief history of Toxicology – from Sumerian drugs to pharmacokinetic analysis of toxicity

There are numerous examples of "drug" usage in ancient times. The first documented evidence of drug receipts is believed to be approximately 5000 years old, on a Sumerian clay slab [1]. In contrast to the long history of using substances from plants for therapeutic purposes, it was only a couple of hundred years ago that people realized the hazards of these substances. This insight can be expressed as "All substances are poisons; there is none which is not a poison. The right dose differentiates a poison and a remedy" [2]. While Paracelsus (1493-1541) had this key insight, the boundary between poison and remedy is hazy. The toxicity of individual chemicals is indeed a complex feature which itself depends on several factors, such as dose, chemistry, individual genetic make-up and exposure to environmental conditions, which all play key roles, to different degrees, in determining susceptibility to disease and adverse drug responses. In modern times it has become increasingly evident that it is not the case that each medicine works equally well, as regards both efficacy and safety, in individuals in a population—hence the rationale behind the idea of personalized medicine [3]. Following the work of Paracelsus, Mathieu Orfila (1787-1853) first described specific organ damage caused by toxins. Toxicity studies of individual substances using animals began in 1920. J.W. Trevan proposed the concept of a 50% lethal dose (LD50), defining the lethal dose of individual chemicals. As a new subject, the field of toxicology slowly developed until the occurrence of the thalidomide disaster in the early 1960s, one of the gloomiest episodes in pharmaceutical history. The drug was approved as a mild sleeping pill with a good safety profile and beneficial effects on morning sickness in pregnant women. However, this caused thousands of babies worldwide to be born with malformed limbs in less than 4 years. Since then, all regulatory agencies have made it obligatory to report the toxicity profiles of Investigational New Drugs (IND). In the late 1980s, the Organization for Economic Co-operation and Development (OECD) and the International Conference on Harmonization (ICH) brought out the guidelines for the toxicity testing of pharmaceutical substances, which are still in use, supplemented with occasional amendments. In the context of regulatory guidelines, the lowest dose able to induce adverse effects (LOAEL) and the highest dose without observable adverse effects (NOAEL) must be tested to extrapolate the derived no-effect level (DNEL), which is more useful in defining the appropriate dose in clinical trials. Other conventional toxicity testing includes repeated dose toxicity testing, carcinogenicity testing, one-generation reproduction toxicity testing, and two-generation reproduction toxicity testing, et al. These depend on the formulation and indication of the drug. The toxicity testing of pharmaceuticals depends strongly on different animal models. Not surprisingly, such an evaluation is expensive (reported to cost more than $3B per year), time-consuming (two-generation reproduction toxicity testing takes around 2 years), suffers from low throughput, and in some cases raises ethical concerns relating to animal welfare [3]. The low throughput of toxicity testing methods has serious consequences for public health, as 86% of chemicals (not limited to drugs) currently on the market lack the necessary toxicity data [4, 5]. The most controversial issue is the translational efficiency of those compounds being tested in humans [6]. No doubt, the current toxicity model is not optimal, motivating both regulatory authorities and pharmaceutical companies to promote innovative alternatives to limit the use of animals and to better assess the risk of drug candidates as early as possible. In 2003, an EPA report proposed a computational toxicology research agenda promising several advantages, including prioritizing candidates and developing predictive models for quantitative risk

assessment. Yet the use of computational methods to predict toxicity has a history in toxicology. In 1962, Hansch et al. developed a Quantitative Structure-Activity Relationship (QSAR) model to estimate the concentration of chemicals using the octanol/water partition and the Hammett constant, which laid the foundation for *in silico* toxicity prediction [7]. Numerous tools were developed to predict carcinogenicity, mutagenicity, and developmental toxicity using pre-built QSAR models such as TopKat and METEOR, most of which have been modified and are currently deployed in academia and the pharmaceutical industry [8] . QSAR models provide a wide range of complexity for toxic endpoints, given flexible feature selection, i.e. qualitative and quantitative toxicity plus molecular descriptors can be used. Yet, QSARs require a large dataset to produce robust statistics, which makes the framework less useful in applications where data is limited. Benezra [9] used structural alerts (SAs) (also called toxicophores/toxic fragments) for skin sensitization in 1982, which was more practicable and economical with the low throughput experimental technologies available at the time. SA based models flourished in toxicity prediction in almost all types of toxic endpoint [10, 11]. Several expert systems are available for toxicity prediction based on pre-built rules and SAs, e.g. HazardExpert, Oncologic Cancer Expert System (OCES), Toxtree, et al.[12–14]. These models are limited to producing qualitative binary output, i.e. toxic or non-toxic. Chemical similarity cluster methods take into account the structural similarity of chemicals, physiochemical features, ADME and mechanisms of action (MoA), which in turn can provide qualitative or quantitative predictions depending on the toxicity endpoint [15]. Multiple tools implement this approach, such as AMBIT, DSSTox and Toxmatch, with applications including prediction of environmental risk, reproductive toxicity, skin sensitization and so on [16–18]. The statistically-derived rule-based approaches mentioned above share a common limitation, namely, lack of biological insights into the mechanistic basis of toxicity. Analogous to pharmacokinetics/pharmacodynamics features indicating the mutual interaction of recipient and chemicals, toxicokinetics/toxicodynamics analysis selects the toxic response related to the chemical concentration *in vivo*. Importantly, measurement of the internal doses rather than administered doses and key metabolites provide a more accurate relationship to the response. In addition, it is a well-developed practice to extrapolate between various administration routes, as different species use non-identical PK/PD and ADME. However, the toxicity pathway and the MoA can only be defined with expert knowledge [19–21]. Drug toxicity is a complex response occurring at system, tissue, cellular and molecular levels. Classic toxicity testing and prediction methods, using either animals or *in silico* chemicals, similarity based or PK/PD based models, simplified complexity and left the mechanistic understanding of the chemical-induced toxicity pathways out of consideration. In 2007, the NRC released the report *Toxicity testing in the 21$^{st}$ century: A Vision and a Strategy*, in which it addressed future directions that would take complexity and toxicity pathways into account [22].

**From Systems Biology to Systems Toxicology**

The revolution in biomedical science in the post genome era has made it feasible to study the effects of chemicals using cells, cellular components and tissues, preferably of human origin. High-throughput assay technologies, bioinformatics and systems biology have significantly empowered scientists to decipher how molecular components, different cells or tissues cooperate to carry out normal physiological functions that are key to maintaining health [23, 24]. Three high-throughput assays developed in recent decades have provided major impetus to the field of toxicology: omics technologies, image techniques, and automated robotic platform techniques.

The platforms enable testing of huge numbers of chemicals in a high-throughput number of samples under standardized conditions. Omics technologies collect the molecular responses to a substance while image methods decode the phenotypical and functional change of cells, organs or organisms in response to exposure to a compound. Together, these three technologies allow researchers to characterize toxicity rapidly at affordable cost [25–27]. As an interdisciplinary field of science, bioinformatics combines computer science, statistics, mathematics, and engineering to analyze and interpret biological data, and serves as a key tool with which to decode the enormous quantum of data generated with high-throughput assays [28]. Since 2000, Systems Biology had been used widely to "understand biology at the system level" using computational and mathematical modeling of complex biological systems [29]. The emergence of systems toxicology can be characterized as the integration of classical toxicology with the quantitative analysis of large networks of molecular and functional changes occurring across multiple levels of biological organization. This is in essence a holistic approach to deciphering the impact of environmental agents (chemicals, complex mixtures, occupational exposures, physical agents, biological agents, and lifestyle factors) on complex biological systems using an engineering approach applied to toxicological research [30]. Systems toxicology is rooted in the on-going revolution in biology and biotechnology, and is founded on the premise that morphological and functional changes in cellular, tissue, organ, and organism levels are caused by and cause changes at the omics level. One example is the Human Toxome project launched by NIH/DDD that is intended to test the strategies that combine omics data and computational models, aiming to develop a common, community accessible framework [31]. Another is Tox-21c, which focuses on toxicity pathways, mechanisms/modes of action, and adverse outcome pathways (AOP) in humans. Tox-21c largely overlaps with 3Rs (replace, reduce, and refine) proposed half a century ago [32, 33]. The Systems Toxicology computational challenge, sbv IMPROVER computational challenge, used crowd resourcing to demonstrate that gene expression data from blood cells are sufficiently informative to predict response to smoking in humans and across species translation [34].

Yet, a comprehensive understanding of the mechanisms of drug toxicity in specific cases requires the integration of different data modalities, from changes at the genomic, proteomic, and metabolomics level across several scales of cellular organization. In contrast to classical approaches, systems toxicology resides at the intersection of systems biology and toxicology where chemistry incorporates mechanisms into the predictive framework [35]. To understand how this complex interaction system in cells and tissues leads to toxicity requires the integration of two disciplines that have been increasingly useful in biomedical research: "Systems Biology" and "Quantitative Pharmacology". In systems biology, a system is generally described as a set of nodes (vertices) connected by edges describing functional interactions. These edges can represent physical interactions, functional interactions, and connections between data across several scales. Similarly, in systems toxicology biological networks are the basis for the prediction of drug action in complex biological systems[36].

Systems toxicology models contain expressions that characterize functional interactions within a biological network, which are very useful when drugs act at multiple targets in the network or when homeostatic feedback mechanisms are operative[37]. Therefore, these models are particularly useful in describing complex patterns of drug action such as synergies between different drugs. Although systems toxicology is still in its infancy, it has tremendous potential to change the way we approach biomedical research. It represents a movement beyond a traditional study-centric approach towards a continuous quantitative integration of data across studies and the different phases of drug development. Network-based approaches offer a wide range of possibilities for deciphering and possibly for understanding the complexity of human disease, thereby providing new tools with which to develop novel drugs. Here we review some current efforts and recent methods through the lens of quantitative systems pharmacology (QSP).

**Examples of Predictive Systems Toxicology**

The general notion of a network-based approach rests upon the ambition to connect several entities across the molecular, cellular pathways, organs and systems to facilitate the prediction of the effect of a drug candidate or any kind of perturbation on biological outcomes of interest [38, 39]. The way in which one defines or infers a network from data is the main determining factor of the degree of reliability and applicability of network analysis in drug design. It is crucial to have a clear definition of network nodes early on, edges and edge weights in the specific application case, and in that context to consider data quality and refinements of the data based on genetic variability, aging, environmental effects. Different types of networks such as networks of chemical compounds, signaling networks, gene-gene interaction networks, protein-protein interaction (PPI) networks or metabolic networks and disease networks can be (and have been) used in QSP models and methods [40] . Following the work on inferring a network comes the analysis of the network and its properties. In the last step, the result of analysis needs to be converted to a series of actionable hypotheses, which then need to be tested and validated or refuted (see Fig1).

Drug–target interaction is the first and most common type of network analysis that has been used in QSP models. Interactions between drugs and targets can facilitate the process of drug discovery by deciphering a drug's mechanism of action, thereby assisting researchers seeking new targets for an old (FDA approved) drug as well as new drug candidates for a known target [41–45]. The main source of information in reconstruction of the Drug- Target interaction network (DTN) is the Drug Bank, which is one of the major publicly available integrated sources of drugs and targets. It is a highly comprehensive database combining chemical properties and detailed clinical information about drugs and their targets. It also provides drug-related data feeds for well-known databases such as Uniprot, PubChem, PDB and KEGG [46, 47].

In spite of the fact that mining drug-target interaction data is increasing at an amazing rate [42], drug-target interaction data currently available from public sources are largely incomplete and biased toward targets of common therapeutic interest [48–50]. Biochemical experiments or *in vitro* methods for finding drug–target interaction are costly and time-consuming. An attempt to address the issue of data completeness of drug-target interaction involves using *in silico* methods [51]. For example, docking simulations are extensively used in pharmacology. AutoDock [52] is one of the most complete suites of free open–source software for the computational docking and virtual screening of small molecules to macromolecular receptors. Xie et al. identified drug off-targets by docking the drug into protein binding pockets similar to those of its primary target, followed by mapping the proteins with the best docking scores to known biological pathways, thus predicting potential side effects[53].Classically, the process starts with a target of known three-dimensional structure, and docking is used to predict the bound conformation and binding energy. In most cases, the three-dimensional structure of a target is needed to compute the binding of each drug candidate to the target, which for many targets are still unavailable [54–56]. Wallach et al. have developed a method to mitigate the impact of this important limitation. They utilize a dataset where there is a pairing of drugs with their observed adverse drug reactions (ADRs), the protein structure database and *in silico* virtual docking to identify putative protein targets for each drug and search for correlated pairs of side effects and biological pathways [57]. Another challenge when performing docking simulation is that it is computationally expensive and most of the methods must simplify the problem to make the computation feasible. The reduction of conformational space by imposing limitations on the system, such as fixed bond angles and lengths in the ligand or a simplified scoring function such as those based on empirical free energies of binding to score poses quickly at each step of the conformation search, are the most common short-cuts that are currently used in the field [52, 58].

In a more recent effort, machine-learning approaches have been used for larger-scale predictions of drug–target interactions. The new interactions between drugs and targets can lead to potential insights on previously unidentified side effects for a particular drug. This idea is the basis of another category of systems toxicology methods. Machine learning-based methods mostly use structural and chemical descriptors of drugs and sequences of targets, similarity matrix or (and) any other pharmacological information about drugs as input. Then they use any machine learning method, such as support vector machines (SVMs) or kernel regression, for predicting the drug–target interactions [59–63]. Cobanoglu et al. used the known interactions in the Drug Bank in the form of a bipartite network to train a model that represents each drug and target as a vector of latent variables and assigns weights to drug-target interactions using probabilistic matrix factorization [64]. Approaches that use similarity scores as input are more promising than other approaches [41].

In general, the use of machine-learning algorithms is one of most promising approaches to extracting knowledge from big data using a data-driven framework. However, the performance of machine-learning algorithms relies heavily on data representations called features, and identifying which features are more appropriate for the given task is very difficult. Deep Learning has recently emerged as a promising technique where the features do not need to be hand-crafted *a priori*. Recent success has been accomplished thanks to the availability of fast computations, massive (labeled) datasets and sophisticated algorithms [65]. Machine learning using deep learning is defined by neural networks with multiple hidden layers. Each layer basically constructs a feature from the preceding layers [66]. The training process allows layers deeper in the network to contribute to the refinement of earlier layers. For this reason, these algorithms can automatically engineer or discover features that are suitable for representing the data at hand. When sufficient data are available, these methods construct features attuned to a specific problem and combine those features into a predictor [67]. Deep learning algorithms have shown promise in fields as diverse as high-energy physics[68] , dermatology[69], and translation[70]. DEEPtox is one of the first methods using Deep Learning for computational toxicity prediction [65]. DeepTox normalizes the chemical representations of the compounds and computes a large number of chemical descriptors that are used as input in machine learning methods. As a next step, DeepTox trains several models, evaluates them, and combines the best of them into ensembles. Finally, DeepTox predicts the toxicity of new compounds. In DEEPTox SVMs, random forests, and elastic nets are used for cross-checking, supplementing the Deep Learning models, and for ensemble learning to complement Deep Neural Networks (DNNs). The networks consist of multiple layers of rectified linear units (ReLUs) to enforce sparse representations and counteract the appearance of a vanishing gradient. ReLUs are followed by a final layer of sigmoid output units, one for each task. One output unit is used for single-task learning. Stochastic gradient descent learning has been used to train the DNNs, and both dropout and L2 weight decay were implemented for the DNNs in the DeepTox pipeline for regularizing learning and avoiding overfitting. Of note is the fact that DEEPtox outperformed many other computational approaches like naive Bayes, support vector machines, and random forests in toxicity prediction of 12,000 environmental chemicals.

The output of all the above-mentioned methods is a DTN, an undirected bipartite network composed of two sets of nodes, drugs and targets. DTN have a complex topology that reflects the inherently rich polypharmacology of drugs (also known as drug repurposing) [51]. The analysis of DTN has recently emerged as an effective means to study targets and to identify new targets for known drugs. In one of the very first attempts, Ma'ayan et al. [71] reconstructed such a bipartite network, and the nodes have been connected if there is an association between a drug and a target on the basis of data from the Drug Bank. They report several classes of proteins as better targets for drugs based on network statistics and gene ontology. A decade later, Lin et al. [72] have followed the same approach to studying the drug–target interaction and could characterize the drug–target relations of different kinds of drugs. They showed that the number of multi-target new molecular entities (NME) has increased over the years, but less than single-target NMEs. In both these cases and several other cases in the literature, it has proven useful to analyze the general structure of a network in order to extract new knowledge facilitating the classification of drugs and/or their targets. Structural (graphical) analysis of a network provides insights into the organization and topology of the DTN and targets for hypothesis generation and experimental testing. As a rule this is performed through computation and analysis of network parameters–parameters that quantify different aspects of the network's internal structure, such as parameters measuring centrality, a node or more global parameters such as modularity index, network density, network entropy or network diameter [73]. Several methods have been developed and applied based on network topology, graph theory, and cluster analysis (see [8] for a recent review). Methods based on the similarity of networks is another set of techniques that have been used to uncover novel target or disease-specific changes [74, 75]. A wide range of similarity measures have been used in the literature, ranging from intuitive measures such as the number of edge changes required to get one network from another or the comparison of the top-k nodes to the more complicated ones, such as using an ensemble of different model networks, and the distribution of the best-fitting ensemble. However, it should be kept in mind that the fundamental question of checking whether two given networks have the same structure, network comparison, is computationally expensive, and despite extensive progress in the field, it remains one of the greatest challenges in the field. For example, it is still not known whether graph isomorphism is polynomial solvable or whether it is NP-complete. Therefore most of the current methods in the network comparison field are heuristic, which in turn may affect the outcome strongly, depending on which kind of prior biases exist in the particular method.

All interactions, from protein-protein interactions (PPI) to gene expression and pathways, are useful in the quest to understand the mechanism(s) of interaction between drugs and complex diseases. Remez et al. used predicted drug−protein interactions obtained with a CT-link in combination gene expression data to obtain a projected anatomical profile of a drug and use it for connecting *in vitro* assays with *in vivo* outcomes and predict potential in *in vivo* organ toxicities [76, 77].

Kuhn et al. used a network based on drug–target interaction data and drug–ADR interaction data to systematically predict and characterize proteins that cause drug side effects. They integrated phenotypic data obtained during clinical trials with known drug–target relations to identify overrepresented protein–side effect combinations [78]. Their networks have three types of nodes: drugs, targets, and side effects, and links are identified side effect causality predictors. The authors considered overrepresented protein-side effect pairs, and hypothesized that such overrepresentation could be indicative of causality. Their approach can make predictions for proteins that are the targets of a certain number of drugs. In this context, Yildirim et al. used a bipartite graph composed of FDA-approved drugs and target proteins in the context of cellular and disease networks and quantitatively demonstrated an overabundance of 'follow-on' drugs[79]. The authors overlaid the drug-protein network with a network of physical PPI. They demonstrated a significant increase in the number of interacting proteins as compared to the average in the PPI network. They used the distance between drugs and a drug target and the corresponding disease to show that most drug targets are not closer to the disease genes in the protein interaction network than a randomly selected group of proteins.

Similarly, several other approaches have been developed based on the notion of expanded drug-target interactions, combined with protein-protein interactions data, in order to develop a network-based pharmacology that could better explain the drug-phenotype relationship, and this approach has been used to predict novel targets and drug repositioning [80–85]. For example, Guney et al. in [86]integrated protein–protein interaction, drug-disease association and drug-target association data. They analyzed the topological characteristics of drug targets with respect to disease proteins and showed that for a drug to be effective against a disease, it had to target proteins within or in the immediate vicinity of the corresponding disease module. Such approaches were also considered for issues related to drug safety and side effects. Cami et al. constructed a network representation of drug-ADR associations for approximately 800 drugs and ADRs and pharmacological information for toxicity prediction. They exploited network structure to predict likely unknown adverse events using a trained logistic regression model [87]. Berger et al. used PPI networks to predict and identify drugs that likely cause Long QT Syndrome based on both a direct drug-target interaction and separate neighborhood [88] .

Complementary to protein-protein interactions, transcriptomic data and gene expression differentiation have been used in drug discovery and safety [88–93]. For example, Gottlieb et al. introduced a method for inferring drug-specific pathways [89]. They connect known drug associated genes over protein, metabolic and transcriptional interaction networks while preferring high confidence interactions participating in curated cellular processes. They use their computed pathways to suggest novel drug repositioning opportunities, gene-side effect associations, and gene-drug interactions. Huang et al. developed a new metric to measure the strength of network connection between drug targets to predict the pharmacodynamics of drug-drug interactions [92, 93].

For the purpose of predicting drug toxicity, in most cases we require a collection of experimental data reflecting molecular changes in the context of quantifiable cellular changes across different biological scales that are linked to toxicity at the body level [35]. So in addition to all the above-mentioned data, systems toxicology depends strongly on the quality and scope of databases annotating side effects (SIDER) and drug-induced differential gene expression, or a combination thereof [94–97].

As an example, Lounkine et al. developed an association metric asking how to prioritize those new off-targets that explained side effects better than any known target of a given drug, thereby creating a drug–target–adverse drug reaction network [43]. Network-based approaches allow the generation of hypotheses about drug-target-phenotype-side effect associations but currently available interaction data are incomplete and the available parts are often non-homogeneous and biased. This situation results in the fact that the conclusions of such studies strongly depend not only on the quality, but importantly, on the degree of completeness of the data [98]. The other relevant point is that most of the suggested approaches in QSP are largely based upon the analysis of the structure of a network or on comparison of networks, while it has been shown[99] that network dynamics, the study of temporal changes in network structures or describing changes of phenotypes of a complex system in the state-space, is crucial to understanding the complexity of diseases and the action of drugs[39]. In this context Mucha et al. [100] developed the technique of multilayer networks, incorporating different types of nodes and edges, in order to follow the changes in module structure in a system having multiple and different types of edges. Interestingly these methods have also been used to predict drug synergies. However, most of them are limited to estimating target links on the PPI network. The advantage of using a network-based approach lies in that it helps to explain the hidden molecular mechanism of drug synergy from the interactions. Due to their effectiveness, some approaches aiming at identifying synergistic drug combinations are based on the dynamic simulation of specific subnetworks. However, these models relied on a very detailed dynamical model, where the lack of information and the uncertainties involved in their kinetics parameters and lots of artificial constraints often limit the usefulness of the simulations, resulting in the model working only for a few specific pathways. Kiani et al. developed a novel integrative pipeline for systematic exploration of drug combinations as a comprehensive and flexible network-based model in the context of the DREAM challenge, a pipeline called HotPPI. Here they constructed a human protein interaction network from major PPI resources, and included both experimentally validated and computationally predicted interactions. The overall procedure resulted in a vast protein interaction network comprising 15,383 proteins and 337,413 interactions. Next, PPI was filtered based on targets of the DREAM challenge and the top 50 pathways involving these targets (table 1). The filtered PPI network comprises 6000 proteins and 16000 interactions. Molecular data are used to weigh interaction in our PPI. The main goal of HOTPPI is to find the best combination to eliminate cancer cell lines. Therefore any combination that eliminates most interactions in a network can cause network collapse followed by death of cancer cells. Thus the heat diffusion algorithm is used to predict potential synergistic drug combinations by calculating how efficient drugs are in hitting the top 200 selected nodes in a network based on their betweenness score. The Hot PPI is generally applicable to high-throughput experimental data where the challenge is to select a small number of the most promising combinations for further mechanistic studies. Using this score, we could rank all possible combinations in a reasonable amount of time. Interestingly, we learned that we should not include too many details (i.e. features or molecular components) in our network descriptions, since we may shift our description from optimal towards the 'knowledge of everything,' with the precision of the method dropping drastically as a result. This underscores the importance and challenge of pruning a large, but for the given application reasonable number of features to include in the network model.

| Pathway maps | pValue |
| --- | --- |
| Ligand-independent activation of Androgen receptor in Prostate Cancer | 1,203E-15 |
| Cell adhesion_PLAU signaling | 4,362E-14 |
| Development_EGFR signaling pathway | 8,380E-14 |
| Apoptosis and survival_Anti-apoptotic action of Gastrin | 1,451E-13 |
| K-RAS signaling in pancreatic cancer | 1,920E-13 |
| Development_G-CSF signaling | 7,018E-13 |
| Development_Growth factors in regulation of oligodendrocyte precursor cell proliferation | 1,065E-12 |
| Main growth factor signaling cascades in multiple myeloma cells | 3,352E-12 |
| Development_VEGF signaling and activation | 5,644E-12 |
| Main pathways of Schwann cells transformation in neurofibromatosis type 1 | 1,107E-11 |
| Immune response_IL-5 signaling | 1,172E-11 |
| Signal transduction_PTEN pathway | 1,172E-11 |
| Immune response_IL-15 signaling | 1,331E-11 |
| Ovarian cancer (main signaling cascades) | 1,595E-11 |
| Tissue Factor signaling in cancer via PAR1 and PAR2 | 2,309E-11 |
| Development_EPO-induced Jak-STAT pathway | 2,549E-11 |
| Apoptosis and survival_HTR1A signaling | 2,864E-11 |
| Development_GM-CSF signaling | 2,864E-11 |
| Cytoskeleton remodeling_TGF, WNT and cytoskeletal remodeling | 3,189E-11 |
| HBV signaling via protein kinases leading to HCC | 3,373E-11 |
| Development_Delta-type opioid receptor mediated cardioprotection | 4,422E-11 |

| Pathway | p-value |
|---|---|
| Apoptosis and survival_Anti-apoptotic action of membrane-bound ESR1 | 5,750E-11 |
| Translation_Non-genomic (rapid) action of Androgen Receptor | 9,496E-11 |
| Apoptosis and survival_BAD phosphorylation | 1,525E-10 |
| Development_Growth hormone signaling via PI3K/AKT and MAPK cascades | 1,525E-10 |
| Development_Ligand-independent activation of ESR1 and ESR2 | 2,960E-10 |
| Development_Membrane-bound ESR1: interaction with growth factors signaling | 2,960E-10 |
| Development_VEGF signaling via VEGFR2 - generic cascades | 3,413E-10 |
| Role of Tissue factor-induced Thrombin signaling in carcinogenesis | 5,194E-10 |
| Development_CNTF receptor signaling | 7,526E-10 |
| IL-6 signaling in multiple myeloma | 9,699E-10 |
| Some pathways of EMT in cancer cells | 9,699E-10 |
| Development_IGF-1 receptor signaling | 1,164E-09 |
| Development_FGF-family signaling | 1,164E-09 |
| Signal transduction_Additional pathways of NF-kB activation (in the cytoplasm) | 1,391E-09 |
| Development_Growth factors in regulation of oligodendrocyte precursor cell survival | 1,563E-09 |
| Development_Dopamine D2 receptor transactivation of EGFR | 1,746E-09 |
| Aberrant B-Raf signaling in melanoma progression | 1,965E-09 |
| Immune response_TSLP signaling | 2,453E-09 |
| Development_Prolactin receptor signaling | 3,215E-09 |
| Development_VEGF-family signaling | 3,751E-09 |
| Immune response_IL-7 signaling in B lymphocytes | 5,605E-09 |
| Signal transduction_AKT signaling | 5,605E-09 |
| Regulation of Tissue factor signaling in cancer | 5,605E-09 |
| Immune response_IL-4 signaling pathway | 6,797E-09 |
| Influence of smoking on activation of EGFR signaling in lung cancer cells | 6,797E-09 |
| Immune response_TNF-R2 signaling pathways | 8,203E-09 |
| Development_Role of IL-8 in angiogenesis | 9,139E-09 |
| Immune response_IL-4 - antiapoptotic action | 9,803E-09 |

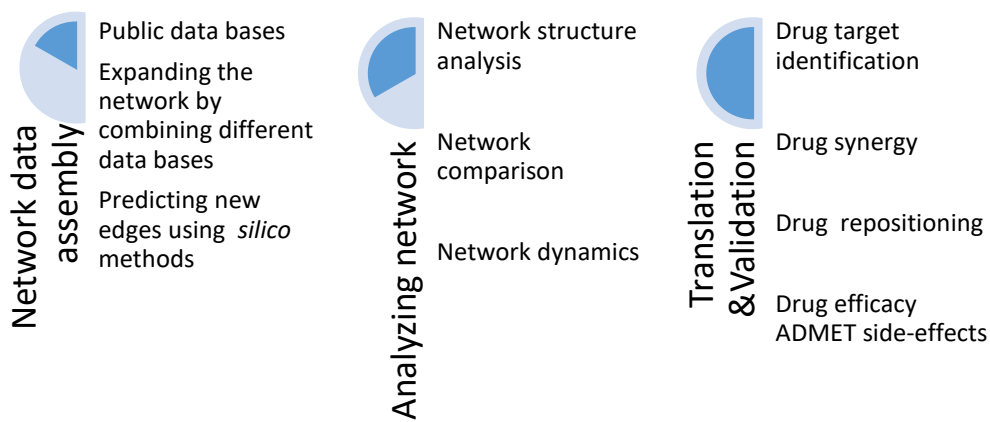

**Fig. 1 overview of predictive system toxicology approaches**

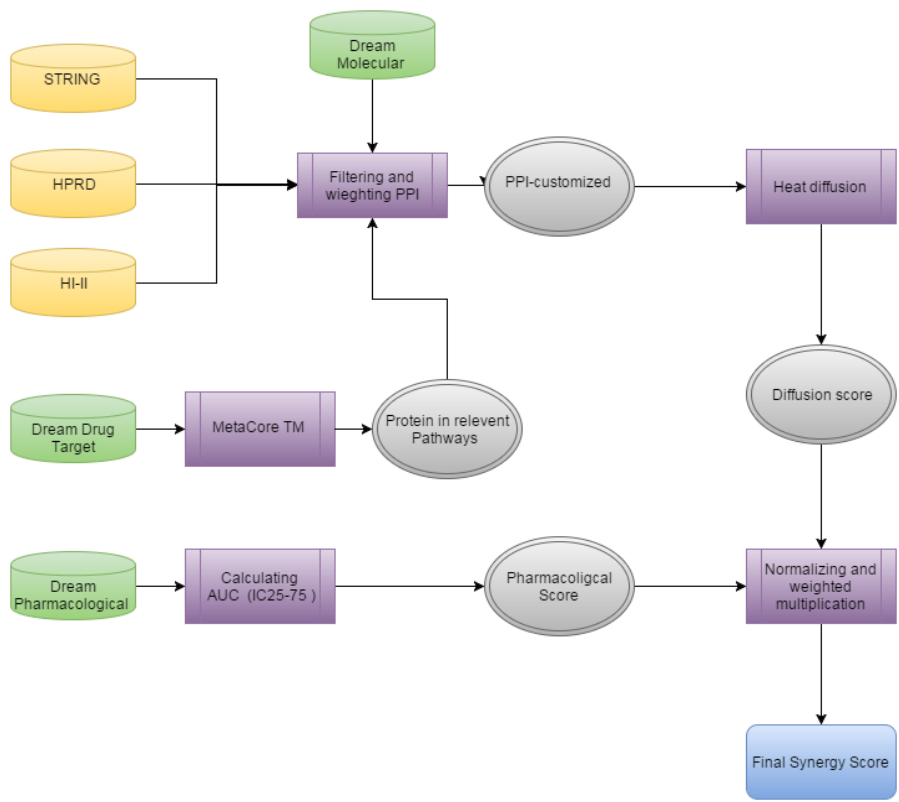

**Fig. 2 Overview of HotPPI approach**

**Information theoretic approach to toxicity**

Both network analysis and pharmacokinetic analysis share a focus and grounding in the physical and functional interactions between molecules within the cell or tissue and the corresponding drugs. Here the overarching aim is not only to predict but also to be able to interpret the mechanism in terms of the underlying biology and chemistry. Since the design of new drugs for new targets is difficult, and the prediction problem is easier from an inference point-of-view, compared to elucidating the mechanisms driving toxicity, complementary approaches are warranted. For example, instead of engineering a drug to target the unique pathways or mutations of a tiny subset of diseases, drug repositioning, such as the one exemplified in the DREAM challenge, involves starting with approved drugs to find combinations that can be used to treat diseases different from the ones they have been designed for, with the advantage that approved drugs can bypass much regulation if correctly controlling for the effects they can have. Thus prediction and simulation are key. This means that the whole field has to move towards causal modeling and functional inference rather than traditional statistical classification (e.g. Tanimoto coefficients) or computational simulation based on classical geometric approaches (e.g. distance between molecules, grid-based docking). To this end, information indexes can facilitate the characterization of drugs by the combinatorial and structural properties shared with or at a remove from the structural properties of the targets, because just as for any molecule, structure means function. Then all these approaches can contribute to determining drug function based on the fact that structurally similar molecules usually have similar properties (known as "neighborhood behavior"). For example, statins are associated with the heart and cholesterol, while morphine, codeine and heroin share structural properties and effects. However, algorithmic information-theoretic approaches based on both classical information and computability theory introduce predictive causal models that go beyond statistical similarities and can find, in principle, similar mechanisms shared by sets of drugs with respect to targets and functions.

It is not difficult to see that complementary regions between drug and target will have a similar classical and algorithmic information content, because the structure of one is the complement of the other. Another advantage is that these measures are parameter-free and thus require no training, even though they can complement and guide machine learning approaches [101, 102]. Because drug docking is not invariant to, e.g., scaling factors, but information theoretic measures are, they may fail to characterize the positive or negative docking properties of a drug. While coarse-graining techniques may be introduced, algorithmic complexity has the advantage of being able to account for scaling effects. The basic idea is the likelihood of a drug being causally generated by a mechanistic model (an algorithm). This is, in general, hard if not impossible to find (the problem is uncomputable), but approximations are possible and new numerical methods have been advanced complementary to statistical and lossless compression approaches that cannot or are very limited at accounting for causation. Drugs, and molecules in general, can be represented in many ways (see Fig. 3a,b,c), some of which are natural networks or networks representing properties of the molecules. Most of these representations are lossless representations, meaning that they can reconstruct the primary representation of the molecule that they encode, e.g., the simplified molecular-input line-entry system or SMILES. The SMILES of a molecule is a string obtained by printing the symbol nodes encountered in a depth-first tree traversal of a chemical graph. SMILES can be converted back (almost) uniquely to the 2-dimensional representation of a drug.

Fig. 3 shows some of these network (a,b) and 2-dimensional representations (c), together with 2 figures (d,e) plotting 3 information-theoretic indexes, two classical and one algorithmic based on the drugs' contact networks. While the 2 classical indexes are the ones most correlated, as one is extracted from the other with the additional information of the sequence valence, the length of the algorithmic complexity (Z-axis) represents the complexity of a hypothesized model producing the contact network.

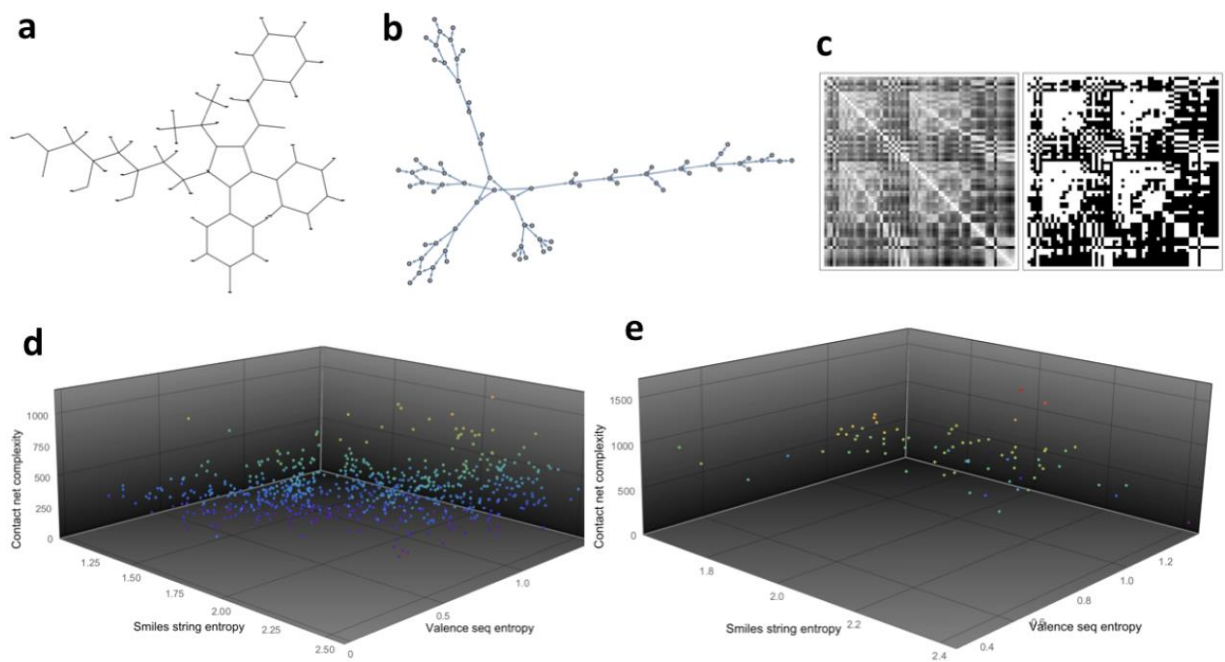

**Fig. 3 Drug profiling by (algorithmic) information indexes: (a)** The molecular (chemical) graph of Atorvastatin ($C_{33}H_{35}FN_2O_5$), a member of the drug class known as statins used primarily as a lipid-lowering agent for prevention associated with treatment of cardiovascular diseases. **(b)** In a molecular network geographical coordinates and shapes are no longer important, but rather their topology (which element is connected to which other), which can be built upon **(c)** the molecular contact map where grey scale (left matrix) indicates proximity between each element that can be binarized (right) using a cut-off value based on the grey scale median. **(d)** Algorithmic information landscape of more than 4000 drugs from the DrugBank (extracted from the Wolfram Language) constructed by taking the entropy of their SMILES codes, the valence sequences of each of the elements in their formula (from SMILES), given the importance they have for bonding **(e)** Algorithmic information landscape of the drugs involved in the DREAM challenge. Color is determined by the 'contact map complexity'; the less complex (the shorter the length of the algorithm generating it) the closer to blue, the longer (more algorithmic-random) the closer to red.

**Concluding remarks**

Here we have reviewed different attempts to predict toxicity from observations (i.e. the Sumerian) to classical pharmacokinetic, advancing to recent integrative systems oriented approaches taking more data into account. These systems approaches resort to performing advanced statistical analytical data processing complemented with machine learning techniques to generate paradigms attempting not only to predict toxicity but also to identify (molecular) mechanisms of toxicity. Information theoretic approaches can be situated in between, as they are as a rule less dependent upon biochemical representations in their problem formulation, while the ones presented here also aim for causal understanding of toxicity in addition to targeting prediction.

In a broader perspective, there are several immediate challenges where we need more work. These include which features to include when predicting toxicity? Minimal models may suffer from being less understandable from a mechanistic standpoint, whereas including too many features, as in the dream example above, could hamper the prediction capability of the model. Overall, a systems biology approach extends the feature space compared to classical pharmacokinetics, while an (algorithmic) information approach facilitates predictions in combination, being both scale invariant and parameter free. Hence there is a tension between predicitive capacity and mechanistic interpretability.

Furthermore, overtraining and overfitting in solving high-dimensional and complex nonlinear problems such as toxicity prediction is one of the most common problems of existing machine learning methods. This originates from the need for estimating and optimizing numerous hyper parameters. However, a method such as the relevance vector machine method solves this problem by incorporating Bayesian criteria into the learning process to reduce the irrelevant support vectors of the decision boundary in feature space, thus resulting in a sparser model[103]. Methods such as Random Forest classifiers are another category of successful methods in systems toxicology. They are one of the most robust algorithms and are able to identify the patterns important for the preferred class, even when there is a large imbalance in the class distribution within the training dataset [104]. Inspecting the results of the TOX21 data challenge demonstrates that a hybrid strategy which combines similarity scores for structural fingerprints and molecular descriptors (features) and machine-learning based prediction models can readily improve the accuracies of toxicity prediction [105]. In general, an ensemble model can be effective, since taking into account the prediction of other models can compensate for an incorrect prediction on the part of one of the individual methods. Certainly, each of the systems toxicology methods has intrinsic advantages, limitations, and practical constraints. Moreover, the performance of these methods depends on the structural diversity and representativeness of the molecules in the data set. Therefore, it is quite important to choose the most suitable machine learning method to develop the prediction model for a specific toxicity data set. Finally, the computational cost associated with each method is another practical and important factor determining the usability of a given method.

In conclusion, beyond the above challenges and considerations, the grand remaining challenge is to advance the state-of-the-art towards personalized medicine. This requires patient specific predictions on toxicity, which in turn requires proper stratification of patients with regard to how they respond or not, with or without adverse toxic effects. This most likely requires integration of multiple layers of information as a background upon which an individual has to be characterized/described, while a machinery for toxicity prediction has to be specific enough for a given patient, given the amount of (sparse) patient-specific information. This challenge and perspective will keep the field of data-driven computational toxicology busy.

## ACKNOWLEDGMENTS

We thank the contributors to HotPPI, Dr. Gordon Ball , Dr. David Gomez, Alireza Mazaheri and Dr. Francesco Marabita. This study was supported by the Swedish ResearchCouncil ( J.T. & H. Z. ) and Karolinska Institutet funds (N.K. ). The funders had no role in study design, data collection and analysis, decision to publish, or preparation of the manuscript.